\documentclass[aps,twocolumn,showpacs,amsmath,amssymb,amsfonts]{revtex4}
\usepackage{graphicx}
\usepackage{bm}
\newcommand{\tst}{\textstyle}

\newcommand{\mbf}{\mathbf}

\begin{document}

\author{Zbigniew Idziaszek$^{1,2}$ and Tommaso Calarco$^{1,3}$}
\affiliation{$^1$CNR-Istituto Nazionale per la Fisica della Materia,
BEC-INFM Trento, I-38050 Povo (TN), Italy\\
$^2$Centrum Fizyki Teoretycznej, Polska Akademia Nauk, 02-668
Warsaw, Poland\\
$^3$European Centre for Theoretical Studies in Nuclear Physics and
Related Areas, I-38050 Villazzano (TN), Italy}

\title{Pseudopotential method for higher partial wave scattering}

\begin{abstract}
We present a zero-range pseudopotential applicable for all partial
wave interactions between neutral atoms. For $p$- and $d$-waves we
derive effective pseudopotentials, which are useful for problems
involving anisotropic external potentials. Finally, we consider
two nontrivial applications of the $p$-wave pseudopotential: we
solve analytically the problem of two interacting spin-polarized
fermions confined in a harmonic trap, and analyze the scattering
of $p$-wave interacting particles in a quasi-two-dimensional
system.
\end{abstract}


\maketitle

Modeling of two-body interactions is the basic step in
development of theories of many-body systems. In the ultracold
regime, atomic collisions are dominated by $s$-wave scattering,
and interactions can be accurately modeled by the Fermi-Huang
pseudopotential \cite{Fermi,Huang}. The situation changes,
however, in the presence of scattering resonances, which can
strongly enhance the contribution from higher partial waves. Such
possibility has been demonstrated in recent experiments by
employing Feshbach resonances to tune the interactions of
identical fermions in $p$-wave \cite{Regal,Zhang}. In this
context, the development of pseudopotentials for higher partial
wave scattering is of crucial importance for the theoretical
description of ultracold gases with $l\neq 0$ interactions.

There are several approaches in the literature to derive
pseudopotential valid for all partial waves. The first derivation
comes from Huang and Yang \cite{Huang}. Their pseudopotential,
however, is incorrect with respect to $l>0$ waves, as recently
shown in \cite{Roth}. Several alternatives have been proposed
\cite{Roth,Stock,Omont,Kanjilal,Pricoupenko}, which have specific
limitations. For instance, \cite{Roth} entails no regularization
and is only applicable to mean-field theories; \cite{Stock}
requires knowledge of the wave function in the inner region of the
shell potential, and taking the limit of shell radius going to
zero in the final step of the calculation.

In this Letter we address the problem of interactions in all
partial waves. We correct the original derivation of Huang and
Yang and obtain a comparatively simple pseudopotential. Next, we
derive explicit pseudopotential forms for $p$- and $d$-wave
interactions, that are are very convenient in calculations
involving anisotropic external potentials. We illustrate this by
solving analytically the problem of two identical fermions
confined in an anisotropic harmonic trap. We finally turn to
interactions of atoms in low dimensional systems. We apply our
$p$-wave pseudopotential to analyze the $p$-wave scattering in 
quasi-two-dimensional (Q2D)
system, and show the occurrence of confinement-induced resonances
analogous to $s$-wave scattering \cite{Olshanii,Petrov}, and 
$p$-wave scattering in Q1D \cite{Granger}. Our
analysis is of direct interest for studies of controlled
interactions between tightly confined fermionic atoms, relevant
for applications to quantum information processing.

First, we derive the pseudopotential for interactions in all
partial waves. We start from the Schr\"odinger equation for the
relative motion
\begin{equation}
\label{SchrEq}
\frac{\hbar^2}{2 \mu} (\Delta + k^2) \Psi( \mbf{r}) = V(r) \Psi( \mbf{r}),
\end{equation}
where $k^2 = 2 \mu E/ \hbar^2$, and $\mu$ denotes the reduced
mass. We assume that the potential $V(r)$ is central and has a
finite range. Outside the range of the potential, the wave
function exhibits the following asymptotic behavior
\begin{equation}
\Psi_{a}(\mbf{r}) = \sum_{l=0}^{\infty} \sum_{m=-l}^{l} R_{l}(r) Y_{lm}(\theta,\phi),
\end{equation}
where $Y_{lm}(\theta,\phi)$ are spherical harmonics, the radial
wave functions $R_{l}= C_{lm} [j_l(kr) - \tan \delta_l n_l(k r)]$
are linear combinations of spherical Bessel and Neumann functions
$j_l(r)$ and $n_l(r)$, $C_{lm}$ are coefficients that depend on
the boundary condition for $r \rightarrow \infty$, and the phase
shifts $\delta_l$ are determined by the potential $V(r)$. Our goal
is to replace $V(r)$ by contact potential $V_{ps}(r)$, which acts
only at $\mbf{r}=0$ and gives the same asymptotic function
$\Psi_{a}(\mbf{r})$ as the real potential $V(r)$. Following Huang
and Yang \cite{Huang}, we determine the pseudopotential from
$V_{ps}(\mbf{r}) \Psi_a(\mbf{r}) = \hbar^2/(2 \mu) (\Delta + k^2)
\Psi_a(\mbf{r})$. Since $j_{l}(k r)$ is regular at $\mbf{r}=0$,
the only contribution to the pseudopotential comes from the
behavior of $n_l(k r)$ for small $r$: $n_l(x) \sim -
(2l-1)!!/x^{l+1}$. Hence
\begin{align}
\label{SchrEq2}
V_{ps} (\mbf{r}) \Psi_a(\mbf{r}) = & \frac{\hbar^2}{2 \mu}
\sum_{l=0}^{\infty} \sum_{m=-l}^{l} (2l-1)!! C_{lm} \tan \delta_l
Y_{lm}(\theta,\phi) \nonumber \\
& \times \! \left[ \frac{d^2}{d r^2} + \frac{2}{r} \frac{d}{d r} - \frac{l(l+1)}{r^2} \right]\!
\frac{1}{(kr)^{l+1}}.
\end{align}
To calculate the action of the radial part of the Laplace operator
on the function $1/r^{l+1}$ at $r=0$, one has to resort to the
theory of generalized functions. By applying the Hadamard finite
part regularization of the singular function $1/r^{k}$
\cite{Kanwal,Estrada}, one can prove the following identity
\begin{align}
\left[\frac{d^2}{dr} + \frac{2}{r} \frac{d}{dr} - \frac{l(l+1)}{r^2} \right] &
\frac{1}{r^{l+1}} =
\label{RadLap}
\frac{ (-1)^{l+1} (2l+1)}{l!} \frac{\delta^{(l)}(r)}{r^2},
\end{align}
where $\delta^{(n)}(r)$ denotes the $n$-th derivative of the delta
function. In comparison, Huang and Yang calculate (\ref{RadLap})
by mapping on $\delta(\mbf{r})/r^l$, which leads them to the
incorrect result. In the final step of the derivation we express
the coefficients $C_{lm}$ in terms of the regularization operator
$\partial_r^{2l+1} r^{l+1}$: $C_{lm} = (2l+1)!!/[k^l
(2l+1)!][\partial_r^{2l+1} r^{l+1} R_l(r)]_{\mbf{r}=0}$. In this
way we obtain the following form of the pseudopotential:
\begin{align}
V_{ps}(\mbf{r}) \Psi(\mbf{r}) = & \frac{\hbar^2}{2 \mu} \sum_{l=0}^{\infty}
\frac{(-1)^{l+1}\,[(2l+1)!!]^2}{4 \pi (2l)!\, l!} \frac{\tan \delta_l}{k^{2l+1}}
\frac{ \delta^{(l)}(r)}{r^2}
\nonumber \\
\label{Vps}
& \times  \left[
\frac{\partial^{2l+1}}{\partial {r^\prime}^{2l+1}} {r^\prime}^{l+1}
\!\int \! \! d\Omega^{\prime} P_l(\mbf{n} \! \cdot \!\mbf{n}^\prime)  \Psi(\mbf{r}^\prime)
\right]_{\mbf{r}^\prime=0},
\end{align}
where the angular integral over $\Omega^{\prime}$ acts as a
projection operator on a state with a given quantum number $l$.
Here, $P_l(x)$ is the Legendre polynomial, $\mbf{n}=\mbf{r}/r$,
and $\mbf{n}^\prime=\mbf{r}^\prime/r^\prime$. We note that in the
calculation of the matrix elements of $V_{ps}(\mbf{r})$ the
differentiation of the delta function is equivalent to
differentiation of the function that acts on the l.h.s. of the
pseudopotential, with a proper change of sign. Moreover, for
functions behaving like $r^{l}$ for small $r$, one can substitute
$\delta^{(l)}(r) = (-1)^{l} l! \delta(r)/r^{l}$ and
$\delta(r) = 4 \pi r^2 \delta(\mbf{r})$ which shows that the
$s$-wave component of (\ref{Vps}) is equivalent to Fermi-Huang
pseudopotential: $-2 \pi \hbar^2 \tan \delta_0/(\mu k)
\delta(\mbf{r})$, whereas components with $l>0$ differ from the
pseudopotential of Huang and Yang \cite{Huang} by a prefactor.

Now we derive an alternative form of the pseudopotential, with
projection operators expressed in terms of the differential
operators. Such representation is particularly useful for problems
involving anisotropic external potentials, since in this case the
wave function, containing several components of the angular
momentum, has to be projected on a given partial wave interaction.
Let us first focus on $p$-wave interactions. Expressing radial
derivatives in terms of the gradient operator $\partial_r =
\mbf{n} \! \cdot \! \bm{\nabla}$, where $\bm{\nabla} =
(\partial_x,\partial_y,\partial_z)$, and performing the angular
integration in the projection operator, we obtain the following
form of the pseudopotential for $p$-wave interactions
\begin{equation}
\label{Vp}
V_{p}(\mbf{r}) = \frac{\pi \hbar^2 a_p(k)^3}{\mu}
\stackrel{\leftarrow}{\nabla}
\delta(\mbf{r}) \stackrel{\rightarrow}{\nabla}
r \frac{\partial^3}{\partial r^3} r^2,
\end{equation}
where $a_p$ is the $p$-wave scattering length: $a_p(k)^3 = - \tan
\delta_1(k) /k^3$, and the symbol $\stackrel{\leftarrow}{\nabla}$
($\stackrel{\rightarrow}{\nabla}$) denotes the gradient operator
that acts to the left (right) of the pseudopotential. To derive
(\ref{Vp}), in the operator acting to the r.h.s. of the
pseudopotential, we introduced an additional multiplication by
$r$: $\partial^3_r r^{2} \rightarrow \partial_r r \partial^3_r
r^{2} = \mbf{n}\! \cdot \!\bm{\nabla} r \partial^{3}_r r^{2}$,
which is equivalent for terms of the order of $r$, giving the 
contribution to the matrix elements.
In this way we preserve the form of the regularization
operator, which is crucial for an exact treatment of the
interacting atoms. We note that the pseudopotential for $p$-wave
interactions in the form containing a scalar product of gradients
has been derived for the first time by Omont \cite{Omont},
however, in \cite{Omont} it contained an incorrect prefactor, and
did not include the regularization operator. We stress that the
presence of the full regularization operator $\partial_r^3 r^2$ on
the r.h.s. of the pseudopotential (\ref{Vp}) is important, since
for anisotropic external potentials the exact wave functions
obtained in the pseudopotential method contains terms behaving
like $x_i/r$ for small $r$. Such terms, for instance, are not
removed by the operator $\bm{\nabla} \partial_r^2 r^2$
\cite{Kanjilal}.

A similar procedure can be repeated for $d$-wave interactions. In
this case we obtain the following pseudopotential
\begin{align}
\label{Vd}
V_{d}(\mbf{r}) = g_d
\sum_{i,j,k,l} & {D_{ijkl}}
 \!\!\!\!\! {\phantom{\Big)}}^{\leftarrow} \!\!
\Big(\partial^2_{x_i x_j}\Big) \delta(\mbf{r})
\Big(\partial^2_{x_k x_l}\Big)^{\!\! \rightarrow}
r^2 \! \frac{\partial^5}{\partial r^5} r^3,
\end{align}
where $D_{ijkl}= \delta_{ik} \delta_{jl} - \frac{1}{3} \delta_{ij}
\delta_{kl}$, $g_d = \pi \hbar^2 a_d(k)^5/(8 \mu)$ and $a_d(k)$ is
the $d$-wave scattering length: $a_d(k)^5 = - \tan
\delta_2(k)/k^5$. As can be easily verified, the pseudopotential
(\ref{Vd}) does not give any contribution for functions exhibiting
$s$- and $p$-wave symmetries, which results from the implicit
projection on $d$-wave contained in (\ref{Vd}).

{\it Two interacting spin-polarized fermions in a harmonic
trap}.--- We start from the integral form of the Schr\"odinger
equation for the relative motion of atoms
\begin{equation}
\label{GEq}
\Psi(\mbf{r}) = \int \! \! d^3r^\prime \, G(\mbf{r},\mbf{r}^\prime) V_p(\mbf{r}^\prime) \Psi(\mbf{r}^\prime),
\end{equation}
where $G(\mbf{r},\mbf{r}^\prime)= \langle \mbf{r}|
(E-\hat{H})^{-1}| \mbf{r}^\prime \rangle$ is the single-particle
Green function, and $\hat{H}$ is the Hamiltonian including the
external potential. For a harmonic potential
$G(\mbf{r},\mbf{r}^\prime)$ can be represented by the following
integral
\begin{align}
\label{Gho}
G(\mbf{r},\mbf{r}^\prime) = - & \frac{(\eta_x \eta_y \eta_z)^{1/2}}{ (2 \pi)^{3/2} d^3 \hbar \omega}\
\int_{0}^{\infty}\!\!\! dt \, e^{E t/(\hbar \omega)} \nonumber \\
& \times \prod_{k} \frac{ \exp\left( \eta_k \frac{x_k x_k^\prime-\frac12(x_k^2+x_k^{\prime 2})
\cosh(t \eta_k)}{d^2 \sinh(t \eta_k)}
\right) }{\sinh(\eta_k t)^{1/2}},
\end{align}
which is convergent for energies smaller than the energy of
zero-point oscillations $E_0=\hbar (\omega_x + \omega_y +
\omega_z)/2$. For $E>E_0$ the Green function can be determined by
analytic continuation of (\ref{Gho}). Here, $d=\sqrt{\hbar/(\mu
\omega)}$, $\eta_k = \omega_k/\omega$, $\omega$ denotes some
reference frequency, and the product in (\ref{Gho}) runs over
$k=x,y,z$.

Inserting the pseudopotential (\ref{Vp}) into (\ref{GEq}), we
obtain a set of linear equations $\sum_l A_{kl}(E) c_l  = c_k$,
where the coefficients $c_{l} = \left[\partial_{x_l} r
\partial_r^{3} r^{2} \Psi(\mbf{r})\right]_{\mbf{r}=0}$, and the
matrix elements are given by $A_{kl}(E) = \pi (a_p/d)^3
[\partial_{x_k} r \partial_r^{3} r^{2} \partial_{x_l^\prime}
G(\mbf{r},\mbf{r}^\prime) ]_{\mbf{r}=\mbf{r}^\prime=0}$. The
energy levels are determined from the secular equation of the
matrix $A(E)-I$. Below we discuss the case of axially symmetric
traps: $\omega_x = \omega_y = \omega_{\perp}=\eta \omega_z$. For
pancake-shape traps with the anisotropy coefficient $\eta=1/n$
where $n$ is some integer, the equations determining the energy
spectrum have the relatively simple form
\begin{align}
\label{EnPan0}
\frac{d_z^3}{a_p^3} = & - \frac{8}{n} \sum_{k=0}^{n-1}
\frac{\Gamma(\frac{k+1/2}{n}-\frac{\cal E}{2}+ \frac 34)}
{\Gamma(\frac{k+1/2}{n}-\frac{\cal E}{2}- \frac 34)}, \quad m=0, \\
\label{EnPan1}
\frac{d_z^3}{a_p^3} = & - \frac{8}{n^2} \sum_{k,l=0}^{n-1}
\frac{\Gamma(\frac{k+l+1}{n}-\frac{\cal E}{2}+ \frac 14)}
{\Gamma(\frac{k+l+1}{n}-\frac{\cal E}{2}- \frac 54)}, \quad m=\pm 1,
\end{align}
where ${\cal E} = E/(\hbar \omega_z)$, $d_z=\sqrt{\hbar/(\mu
\omega_z)}$ and $m$ is the projection of the angular momentum on
the $z$-axis. As it can be easily verified, for the case of
spherically symmetric trap ($\eta=1$), Eqs.
(\ref{EnPan0})-(\ref{EnPan1}) coincide with the result of Refs.
\cite{Stock,Kanjilal}. We note that in the presence of an anisotropic
the degeneracy between states with $m=0$ and
$m=\pm1$ is lifted. We have derived similar closed analytic
formulas for cigar-shape traps with $\eta = n$
\cite{IdziaszekPrep}, however, we do not present them in the
present letter because of their rather complicated form. In
general, when $\eta \neq 1/n$ and $\eta \neq  n$, the energy
levels can be determined from an implicit equation involving the
integral representation (\ref{Gho}) \footnote{For $s$-wave
interactions see \cite{Idziaszek}}.

\begin{figure}
   \includegraphics[width=8.6cm,clip]{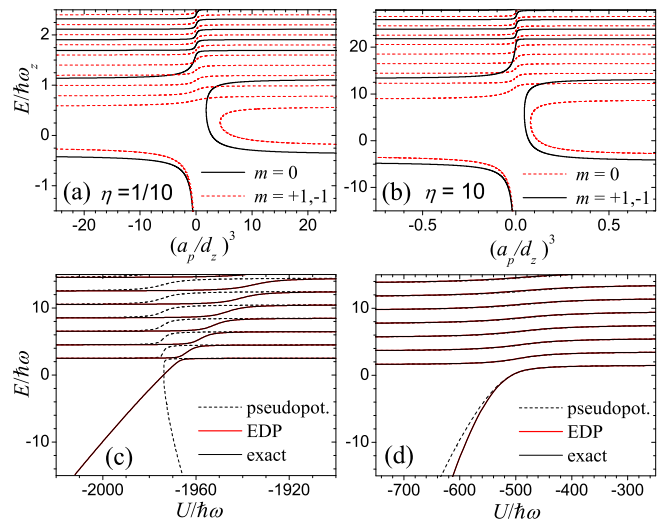}
     \vspace{1mm}
     \caption{
     \label{fig:En} (a)-(b):
     Energy spectrum of the two fermions interacting in $p$-wave as a function of the scattering volume $a_p^3/d^3$,
     with $d = \sqrt{\hbar / \mu \omega_z}$. The two
     particles are confined in an axially symmetric harmonic trap with $\eta = \omega_{\perp}/\omega_z = 0.1$ (a), and
     $\eta = 10$ (b). (c)-(d) Energy spectrum of two atoms interacting via a square-well model potential
     as a function of the well depth $U$. The atoms are
     confined in an isotropic harmonic potential and interact in $p$-wave (c) and $s$-wave (d).
     }
\end{figure}

Figs.~\ref{fig:En}(a) and \ref{fig:En}(b) show the energy levels
in harmonic traps with $\eta=1/10$ and $\eta=10$, for different
values of the quantum number $m$. We observe that for small and
positive values of the scattering volume, the system does not have
eigenstates with energies smaller than the lowest eigenenergy for
noninteracting atoms, This particular behavior results from the
properties of the $p$-wave pseudopotential. In this regime of
energies, however, it is necessary to include in the calculation
the energy-dependence of the pseudopotential. To this end it is
sufficient to insert in the pseudopotential (\ref{Vp}) the energy
dependence of the phase-shift $\delta_1(k)$, and calculate the
energy spectrum in a self-consistent way. This extends the
applicability of the pseudopotential method to the large
scattering lengths and tight confining potentials and allows to
describe the entire molecular spectrum of the realistic potential
\cite{Bolda,Stock}.

To illustrate this procedure we calculate the energy spectrum for
a square-well model interaction, assuming for simplicity
spherically symmetric harmonic trap. Fig.~\ref{fig:En}(c) presents
the energy levels of $p$-wave interacting particles as a function
of the square-well depth $U$ calculated for the square-well radius
$R_0=0.05 d$. The depth $U$ is varied close to the resonance
scattering for $p$-wave, related to the appearance of a bound
state at $U = - 1974 \hbar \omega$. The exact energy levels are
compared with predictions of the pseudopotential method given by
Eq.~(\ref{EnPan0}), and with the energy spectrum calculated in a
self-consistent way from Eq.~(\ref{EnPan0}) with $a_p^3$ replaced
by $a_p(k)^3=-\tan \delta_1(k)/k^3$. For comparison,
Fig.~\ref{fig:En}(d) shows the energy levels of $s$-wave
interacting atoms, given in the pseudopotential approximation by
$d/a_s = 2 \Gamma(\frac 34-\frac{\cal E}{2})/\Gamma(\frac
14-\frac{\cal E}{2})$ \cite{Busch}, where ${\cal E}=E/(\hbar
\omega)$ and $a_s$ is the $s$-wave scattering length.

One observes that self-consistent calculation with
energy-dependent pseudopotential (EDP) provides very accurate
results for the energy spectrum. On the other hand, the ordinary
pseudopotential method fails for large scattering volumes, and for
energies where the bound state of the square well potential
appears. Moreover, its range of applicability decreases for
higher energy levels. For comparison, the ordinary $s$-wave
pseudopotential is incorrect only with respect to deep
bound-states.

This behavior can be explained by analyzing the effective range
expansion: $k^3 \cot \delta_1(k) = -1/a_p^3 - k^2/(2 R^{\ast}) +
{\cal O}(k^4)$, where $R^{\ast}$ can be interpreted as effective
range for $p$-wave (for square-well $R^{\ast} = R_0/3$). The
second term in the expansion can be neglected when $k |a_p| \ll (k
R^{\ast})^{1/3}$, which combined with the condition for the
applicability of the pseudopotential: $k R^{\ast} \ll 1$, gives $k
|a_p| \ll 1$. Therefore, the regime where one can apply the
ordinary, energy-independent pseudopotential for $p$-waves is
quite narrow. This can be compared to the $s$-wave, where the
analogous condition takes form $k |a| \ll (k \tilde{R})^{-1}$,
where $\tilde{R}$ is the effective range for $s$-wave scattering
and $(k \tilde{R})^{-1}$ is large.

{\it Scattering of spin-polarized fermions in Q2D systems.}---
The scattering solution
can be found from the Lippmann-Schwinger equation:
\begin{equation}
\label{LS}
\Psi(\mbf{r}) = \Psi_0(\mbf{r}) +
\int \! \! d^3r^\prime \, G_{+}(\mbf{r},\mbf{r}^\prime) V_p(\mbf{r}^\prime) \Psi(\mbf{r}^\prime),
\end{equation}
where $\Psi_0(\mbf{r})$ represent the wave function of the
incoming particle and $G_{+}(\mbf{r},\mbf{r}^\prime)= \langle
\mbf{r}| (E-\hat{H}+i \epsilon)^{-1}| \mbf{r}^\prime \rangle$. The
Green function $G_{+}(\mbf{r},\mbf{r}^\prime)$ describing the
propagation of outgoing waves, can be determined from
Eq.(\ref{Gho}) by taking the limit $\omega_\perp \rightarrow 0$
and performing the analytic continuation for energies $E>E_0$.

In 2D the scattered wave in the asymptotic regime ($\rho
\rightarrow \infty$) is described by  $f(\phi) e^{i k
\rho}/\sqrt{\rho}$ with the scattering amplitude $f(\phi) = (2 \pi
i k)^{-1/2} \sum_{m=-\infty}^\infty e^{i m \phi} (e^{i 2
\delta_{m}} -1)$, where $\delta_{m}$ are scattering phase-shifts
\cite{LL}. In the regime of energies $\hbar \omega_z /2 < E < 3
\hbar \omega_z /2$ the motion in the $z$ direction is frozen and
sufficiently far from the scattering center it is described by the
ground-state wave function of the harmonic oscillator: $\psi_0(z)
= e^{- z^2/(2 d_z^2)}/\pi^{1/4}$. By solving Eq.~(\ref{LS}) with
the pseudopotential ({\ref{Vp}), we obtain the scattering
solution, which for $p$-wave interactions contains only $m=\pm1$
scattering waves
\begin{equation}
\Psi(\bm{\rho}) \stackrel{\rho \rightarrow \infty}{\sim} \psi_0(z) \left\{
e^{i \mbf{k} \bm{\rho}} -
\frac{4 \cos \phi}{1+ i \cot \delta_1} \frac{e^{i k \rho}}{\sqrt{2\pi i k \rho}}
\right\}.
\end{equation}
\begin{figure}
   \includegraphics[width=6cm,clip]{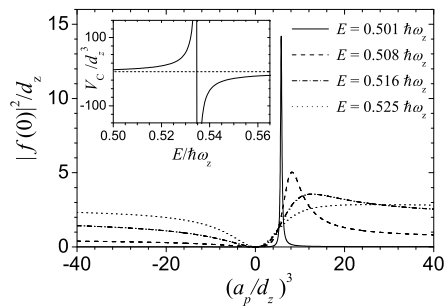}
     \vspace{1mm}
     \caption{
     \label{fig:Fsc}
     Two-dimensional differential cross section in forward direction $|f(0)|^2$ in the scattering of two
     fermions in a quasi-2D system as a function of $p$-wave scattering volume $a_p^3$, for different relative energies
     $E$ of the particles. The inset shows
     the value of the critical volume $V_c$, at which the system exhibits the confinement-induced resonance.
     }
\end{figure}
The scattering phase shift $\delta_1$ is given by
\begin{equation}
\cot \delta_1 = - \frac{2}{3 \pi k^2 d_z^2} \left[ \frac{\sqrt{\pi} d_z^3}{a_p^3} -
{\cal W}\left(\frac{\cal E}{2} - \frac 14\right) \right],
\end{equation}
where ${\cal W}(x)$ is a function taking values of the order of
one \footnote{ ${\cal W}(x)  =  8x + 6x \ln [x/(1-x)] + 4
\sum_{k=0}^{\infty} h_k(x)(2k -1)!!/(2^k k!), \quad h_k(x)  =  2x
- k +(3x-k)(k + {\tst \frac 12}) \ln [(k-x)/(k+1-x)].$}, and
${\cal E} = E /(\hbar \omega_z)$. When $\cot \delta_1 = 0$, the
system exhibits confinement-induced resonance, which occurs at the
scattering volume $V_c({\cal E})= \sqrt{\pi} d_z^3 /{\cal W}({\cal
E}/2 - 1/4)$. At low energies (${\cal E} \rightarrow \frac 12 $)
function ${\cal W}(x)$ has a well defined limit: ${\cal W}(0)
\approx 0.325$, and $V_c(0) \approx 5.4 d_z^3$. This qualitatively
differs from $s$-wave interactions, where $\cot \delta_0$ exhibits
logarithmic behavior for small $k$ \cite{Petrov} and the resonance
condition for $k \rightarrow 0$ depends on the relative kinetic
energy of the scattered particles. Fig.~\ref{fig:Fsc} shows the
dependence of the differential cross section $|f(\phi)|^2$ at
$\phi=0$ on the scattering volume $a_p^3$ for different values of
the energy. We observe that at low energies of the scattered
particles the curve is strongly peaked around $V_c$. For higher
energies the resonance is broader and finally for ${\cal E}
\approx 0.525$ disappears. In this regime of energies the
resonance is present for positive values of the scattering volume,
which can be observed in the inset of Fig.~\ref{fig:Fsc}
presenting $V_c({\cal E})$.

In summary, we presented the zero-range pseudopotential applicable
for all partial wave interactions. For $p$- and $d$-waves we
derived an alternative representation of the pseudopotential, in
which the projection on spherical harmonics is replaced by an
appropriate differential operator. The $p$-wave pseudopotential has
been applied to calculate analytically the spectrum of two
interacting fermions in a harmonic trap, and to study the
scattering of identical fermions in a quasi-two-dimensional
system.

After completing this paper we learned of recent work of Derevianko
\cite{Derevianko} in which the pseudopotential 
equivalent to (\ref{Vps}) is derived.

We thank L.P. Pitaevskii, R. Stock, and G.V. Shlyapnikov for
valuable discussions. T. Calarco acknowledges support from the EC
through contract No. IST-2001-38863.

\end{document}